# 3D printing for teaching and exploration in astronomy for individuals with blindness/visual impairment: textured representations of imagery

Carol A. Christian[a], Antonella Nota[b], Noreen Grice[c], Thomas Madura[d], David Hurd[e]

[a]Space Telescope Science Institute, Baltimore, MD, USA; [b]International Space Science Institute, and Space Telescope Science Institute, Bern, Switzerland; [c]You Can Do Astronomy LLC, New Britain, CT, USA; [d]San Jose State University, San Jose, CA, USA; [e]Penn West Edinboro, Edinboro, PA, USA

## Abstract

Astronomy, a captivating field that draws upon science, mathematics, and engineering, has traditionally relied on visual representations to convey the wonders of the cosmos. While this approach effectively engages the sighted population, the use of imagery can exclude individuals with blindness or visual impairment (B/VI). Astronomical research is incorporated into press releases, media, outreach efforts, and educational systems aimed at enhancing public interest and often skill in science, but visual materials can hamper a population with B/VI. This paper explores the potential of 3D printing as an assistive technology providing an alternative to imagery. We produced textured 3D prints of astronomical research data from the Hubble Space Telescope (HST). Useability assessment of materials is an important phase of production before integration into structured programs, and we used a multi-phased approach in our prior research to create and test appropriate textures for 3D astronomical prints. This paper describes the last step of reviewing our 3D prints through informal useability sessions with diverse individuals. The assessment indicated our 3D prints provide reliable, informative representations of astronomical data appropriate for public use especially for public information, outreach programs, and science education for individuals with BVI.

## Introduction

Astronomy, a popular subject for ages, probes deep questions about the cosmos. Astronomy con-tributes to science, technology, engineering, and math (STEM) education and public understanding of science as well as other aspects of life and culture (National Academies Press, 2001; National Aca-demies Press, 2020; Pompea & Russo, 2020; Rosenberg et al., 2013; Salimpour & Fitzgerald, 2022). Astronomical images and visualization provide the principal sources of information that shape public impressions about the universe, the solar system, and even the space just beyond Earth's atmosphere (Madsen & West, 2003; Peebles et al., 1994; Setti, 2012).

While visual treatments engage the general public, they are not necessarily helpful for those with B/VI (Hasper et al., 2015). Science is usually taught and learned employing visual methods (Wild & Koehler, 2017), but students with blindness/visual impairments (B/VI) have difficulties with such techniques (Chiu, 2020). Resources for those with B/VI are sparse at the national and international levels and the research and development in the field of science education for the blind and visually impaired is slow (Beck-Winchatz, B., & Riccobono, M. A., 2008; Jones et al., 2006; Jones et al., 2009; Supalo et al., 2014; Wild et al., 2022). A by-product of the sparsity of resources is that individuals with (B/VI) may be deterred in considering the possibility of understanding, studying, or even pursuing a career in astronomy, or any STEM field (Arcand et al., 2019).

For many years, we and others have researched the idea that 3D printing, among other assistive technologies and innovative access to information, may hold promise for concept development for anyone with B/VI (Koehler et al., 2018; VanderMolen & Fortuna 2021 and references therein). The escalation of 3D print resources is augmenting existing information materials, other resources, and activities integrated in public communication, education, or outreach programs (c.f., Arcand et al., 2019; Karbowski, 2020, NASALearning, n.d). While 3D printing in education was on the rise in the last decade to support teaching and learning (Ford & Minshall, 2019) it was quite rare in astronomy and other sciences when we initiated our work.

In 2012–2018, we conducted a program to prepare textured 3D print models of astronomical research imagery for communication channels such as press releases (e.g. NASANews, n.d.; HubblesiteNews, n.d.), publications, exhibits, formal education (NASALearning, n.d.), outreach, or individual applications (e.g. home schooling, life-long-learning, etc.) similar to the strategy discussed by Arcand et al. (2019). The program was aimed to provide reliable, understandable, and usable 3D prints that conceptually represent the research data. We designed a multi-phase assessment of our materials in which we produced samples of various 3D printable textures based on past tactile research, including Braille resources (Grice et al., 2015), applying best textures to distinguish the features of an astronomical object, and subsequent scrutiny of the textured prints (Christian et al., 2014; Christian et al., 2015; Grice et al., 2015). This paper discusses the final usability tests of textured 3D prints that model a range of astronomical objects. We assessed functionality, discernibility of individual textures, effectiveness in conveying astronomical concepts, transferability of texture representation across different objects, general impressions of the prints, suggestions for improvement, and any augmentations recommended. As a preview to our result, we found our 3D prints are suitable assessable substitutes for astronomical images.

### Statistics and equity

A glimpse at the extent of individuals affected by B/VI is discussed in depth by a number of organizations and in prior publications. For example, the World Health Organization (2023) cites that individuals with B/VI comprise approximately 2.2 billion people world-wide, providing detailed statistics on impairments and demographics (c.f., also Bourne et al., 2017, and references). National Federation of the Blind (NFB, 2018) provides comprehensive statistics on B/VI: 2.4% of the U.S. population corresponding to slightly less than 8 million individuals. The statistics show visual impairment often worsens as individuals age, aligning with the statistics and modeling from the National Eye Institute (2022), projecting a substantial degradation over the next decade. Surveys by NFB (2018) reveal that nearly one million U.S. school and college-aged individuals have B/VI.

B/VI is recognized as a disability under the Americans with Disabilities Act Title I (ADA National Network, 2019). This legislation mandates equal opportunities and reasonable accommodations for individuals with disabilities in employment and other settings. A lack of suitable resources and support structures for learners with B/VI hinders their potential and creates unnecessary barriers to their scientific exploration intellectual growth, and possible future careers in STEM fields. Arcand et al. (2019) discuss at length the legal structures and the research findings regarding equity standards in the U.S. and world-wide (c.f., Smithsonian, 2019; United Nations, 2006).

## Theoretical background

### Contribution of astronomy to science education and engagement

Astronomy education uniquely engages learners in a multidisciplinary field that integrates physics, chemistry, biology, geology, and mathematics as well as technical disciplines such as computing, optics, mechanics, experimentation, and observational technique (Madsen & West, 2003; National Academies Press, 2001). Astronomy has proven to be important in history, culture, and art (National Academies Press, 2020; Salimpour & Fitzgerald, 2022; Setti, 2012). Pompea and Russo (2020) discussed, in depth, the role of astronomy in science education and teacher professional development as well as the production of educational materials.

Stimulating interest in astronomy can provide a path to engagement in the sciences for everyone (National Academies Press, 2001). To benefit, individuals with B/VI must have alternatives to astronomical images and data (AAAS, 2019; Wild et al., 2022). Such assistive technologies can be used more effectively when used in concert with other proven public communication and education resources as recommended by Wild et al. (2022), Chiu (2020), and others. Tactile resources can be highly beneficial for students with disabilities, promoting improved performance and participation in learning (McCarthy, 2005; Supalo et al., 2016). Bonne et al. (2018) explicitly noted that engaging individuals with B/VI in astronomy exposes them to possibilities of professional scientific work and improves quality of life. Yet, existing resources for learners with B/VI, even Braille and audio materials and especially tactile models, remain in limited availability (Grice, 2007; Grice et al., 2015).

Understanding of scale, spatial relationships between physical objects, and distinct mental models of physical phenomena are all important in astronomy and generally contribute to our ability to reason, solve problems, and understand scientific concepts (Jones et al., 2009; Plummer, 2014; Rule, 2011 and refs therein). Mental model construction, key the learning process, is stimulated with astronomical concepts, and relevant to those with B/VI (Jafri et al., 2017; Grice et al., 2015; Arcand et al., 2010). Jones et al. (2009) found that with practice, people with B/VI can better understand concepts concerning very large and very small spatial scales than their sighted peers. However, until recently, most individuals with B/VI were rarely provided with the tools for enhancement of their conceptualization skills, and even now educators are often at a loss as to the type of tools to use to deliver a stimulating environment for students (Supalo et al., 2016; Wild et al., 2013; Wild et al., 2022).

### Astronomical data

Astronomy is usually considered a visual science, yet celestial objects also produce electromagnetic radiation (EM) outside the visible such as radio, millimeter, Xray and Gamma Ray and, beyond those, through multi-messenger signals (Mészáros et al., 2019) such as gravitational waves (Abbott et al., 2016) and neutrinos. To facilitate communication, data associated with those 'invisible' regimes can be translated to visual materials (Kent, 2017 and related articles in the same volume) or sonic representations (Ferguson, 2016; Harrison et al., 2022). Sound has been effectively used to convey spectral and variable phenomena (e.g. Cooke et al., 2019; Diaz-Merced et al., 2012) and gravitational wave detections (LIGO Multimedia, 2020). Since visual imagery is dominant for communication, we produced representations of that format for accessibility.

Astronomical data are generally obtained with digital instruments on telescopes. For imagery, a series of exposures are taken through a succession of very precisely fabricated filters depending upon the science objective (ESAColor, 2024; HubblesiteColor, 2022; WikipediaAstrObs, 2024).

Scientists conduct analysis on the individual exposures to understand the astrophysics of the object. For public consumption, composite images are created by combining multiple science exposures to enhance detailed astrophysical features of the object in one image, accompanied by descriptions of the science research, technical details, and links to relevant publications (see HubblesiteColor, 2022 and references therein for the process). The color images typically are released in public media (see for example HubblesiteNews, n.d.; and Figure 1 in the Supplemental Material). Such images are the source material for our 3D prints.

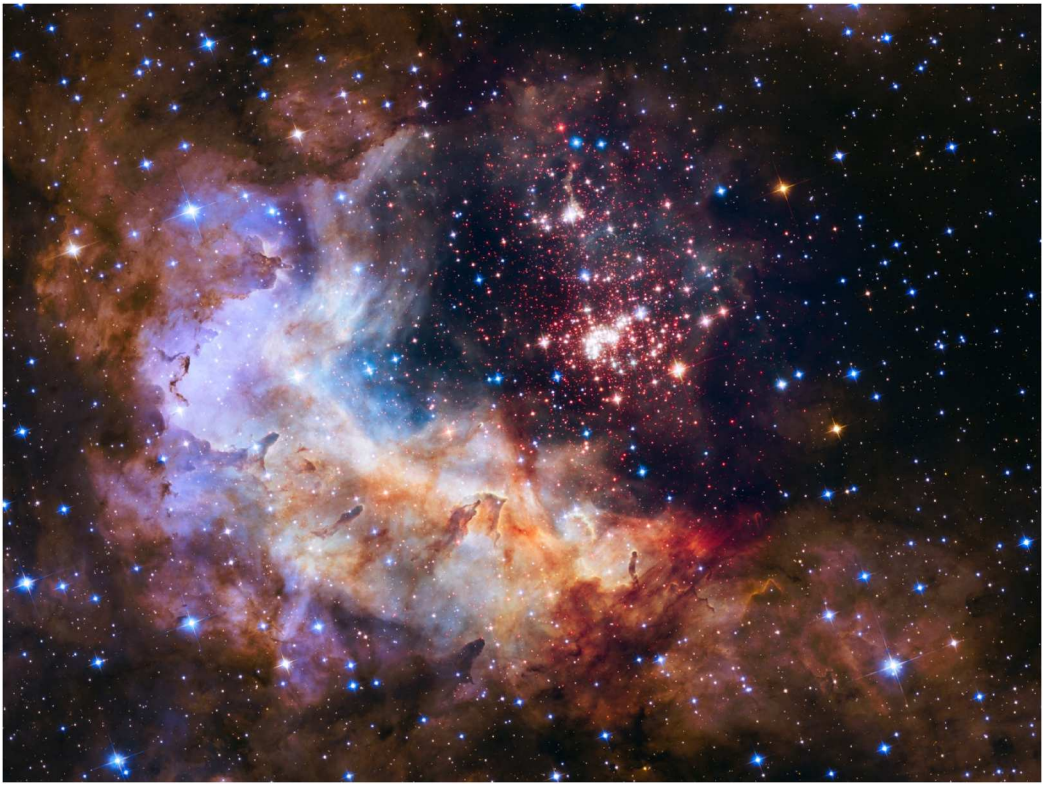

**Figure 1.** HST image of Westerlund 2, a giant cluster containing about 3000 stars surrounded by nebulosity from which the stars were formed. The cluster is actually double, formed serially. The nebula contains gas, dust, and filamentary structures. Red traces hydrogen emission, blue green shows the location of oxygen emission. The research associated with this image of Westerlund 2 is described by Zeidler et al. (2015). Also see Westerlund (1961) and WikipediaW2 (2024). Image credit: NASA and ESA.

The captivating science images are beneficial for authentic research experiences and science curriculum enrichment as discussed by Rector et al. (2015). Also, advanced students can find data, especially from space telescopes, in open archives so that with expert training, they can conduct their own analysis (e.g. NASAData, n.d.). Thus, there are many levels of resources to stimulate an interest in astronomy, however, literature documenting the experience of students with B/VI is scarce (Jones et al., 2006; Koehler et al., 2018; Wild et al., 2013). Studies do suggest that tactile materials, when available, can stimulate students with disabilities and help them perform significantly better in many educational environments (McCarthy, 2005; Supalo et al., 2016).

### Research statement

This study assessed the useability and suitability of uniquely textured 3D prints produced by transforming astronomical image data from Hubble Space Telescope (HST: NASAHST, 2024) into tactile forms using a previously tested production method. We also aimed to understand the potential of 3D prints in enhancing accessibility and curiosity about science.

## Materials

### Technique for creating 3D prints

#### Imagery selection

We concentrated on HST data which has resulted in a wealth of peer-reviewed research. HST is one of the most prolific and public recognizable astrophysical research instruments built, producing more than 1000 refereed papers each year from this one facility (ESAHSTScience, n.d.; WikipediaHST, 2024). HST has been the source of an enormous amount of public information and educational material (NASAHSTInfo, n.d.; NASANews, n.d.; STScIComm, n.d.; Hubblesite, n.d.; WebbTelescope, n.d.). Materials derived from HST science data represent authentic, state-of-the-art science, and include imagery, graphics, videos, text, audio, and computer visualization of astronomical objects. The imagery was our focus for tactical representation. Images of various kinds translated to relief prints have met with some success (Hasper et al., 2015), but our project focused on producing 3D prints with distinguishable textures.

Our data selections were images (HubblesiteNews, n.d.) derived from peer-reviewed research of star clusters in our own Milky Way galaxy and in nearby galaxies (e.g. Zeidler et al., 2015 and subsequent papers; Carlson et al., 2007) and studies of entire galaxies that are more distant (Calzetti, 2015 and subsequent papers). We provide brief additional information about star clusters and galaxies in the Supplemental Material. The first dataset printed were star clusters (WikipediaStarClusters, 2024) which are aggregates of stars that form together from giant clouds of gas and dust that collapse due to gravity. Characterizing star cluster elements, e.g. stars, gas, dust, and compressed filamentary structures, is important for understanding the fundamental components of the rest of the universe.

The second dataset printed were galaxies which are enormous collections of stars, star clusters, gas and dust bound together by gravity (WikipediaGalaxy, 2024). The data include nearby objects of different morphologies (shapes). All of the galaxies chosen have a rich population of star clusters as well as dust and gas, which trace their shapes. Star clusters and their homes, the galaxies, are topics that draw on important astrophysical concepts – atoms, molecules, chemistry, physical processes, velocities, gravity, EM and astrobiology, that is, life in the universe. All are key for understanding the formation and global fate of the universe.

For communications, many astronomical objects are referred to by colloquial names as well as catalog designations. There are hundreds of catalogs from the 1800s through the present (AstroCatalog, 2024). One common catalog is the New General Catalog (NGC, 2024) and other catalogs bear the cataloger's name.

### *Creating star cluster prints*

To make the tactile, textured representation of a star cluster image, the data are loaded into a software module called Astro3D publicly available on GitHub (Bradley et al., 2024; and details in 3DAstroPrints, n.d., and in the Supplemental Material). Star cluster features were assigned specific textures and extensively tested previously (Christian et al., 2015; Grice et al., 2015). The textures preferred by the testers were distinguishable, robust, and intuitive for representing object features. The best textures identified were adopted as a 'standard' for the next phase (this study): stars are represented as cylindrical features with height correlated to brightness, gas assigned a soft 'stipple', a grid of closely spaced bumps, dust assigned a coarser stipple, and filamentary structure assigned ridges of parallel lines. The 3D tactile print of the star clusters Westerlund 2 and NGC602 are exhibited in Figures 2 and 3 (in Supplemental Material) with undulating surfaces determined from the brightness of the image plus the overlaid texture signifying the type of feature at each location (3DAstronomyPrints, n.d.).

### *Creating galaxy prints*

Galaxies are huge collections of billions of stars, gas, and dust. They have a variety of physical sizes, masses, shapes and distributions of their components. Since galaxies have the same fundamental components as star clusters, the textures assigned to star clusters also were appropriate for representing galaxy structure. The exception is that star clusters in galaxy prints are represented not by one cylinder but three together, with the elevation corresponding to the cluster brightness (Christian et al., 2015; Grice et al., 2015). Figures 4, 5, and 6 in the Supplemental Material exhibit the galaxies NGC1566, NGC3344, the Whirlpool Galaxy each with its corresponding print (3DAstronomyPrints, n.d.).

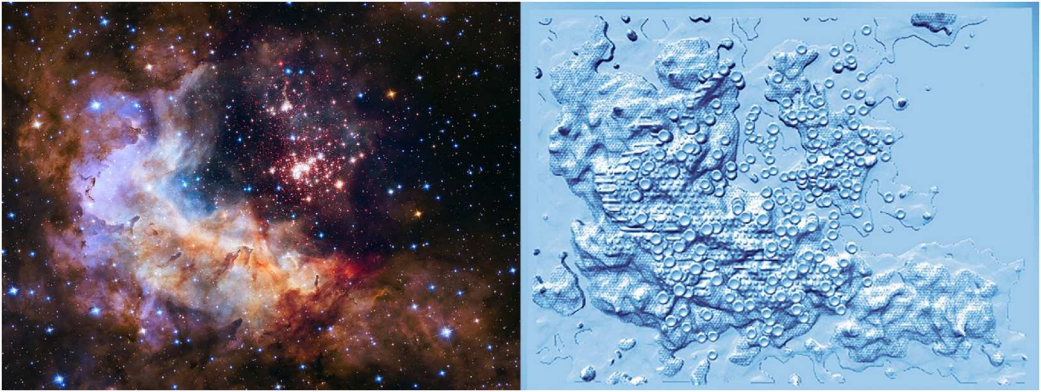

**Figure 2.** Left: HST image of Westerlund 2, Right: 3D Tactile print of the star cluster using our adopted standard textures for features in the cluster described in Figure 1 (3DAstromomyPrints, n.d.). The distance to this cluster is 20 thousand light-years, the diameter about 6–13 light-years, with an age of about 2 million years. Image credit NASA and ESA.

### 3D printing process

Our development project was aimed at creating accessible alternates for imagery for B/VI learners and the public, and all validated 3D print files were to be published online with print guides, and now have been (3DAstronomyPrints, n.d.; NASA3D, n.d.). This allows individuals without much 3D printing experience to produce prints from our files on simple inexpensive 3D printers. The detailed process we used and recommend for printing is described in the project website (3DAstronomyPrints, n.d.) and by Grice et al. (2015) including printer specifications and sizes of prints. A brief summary is presented in the Supplemental Material.

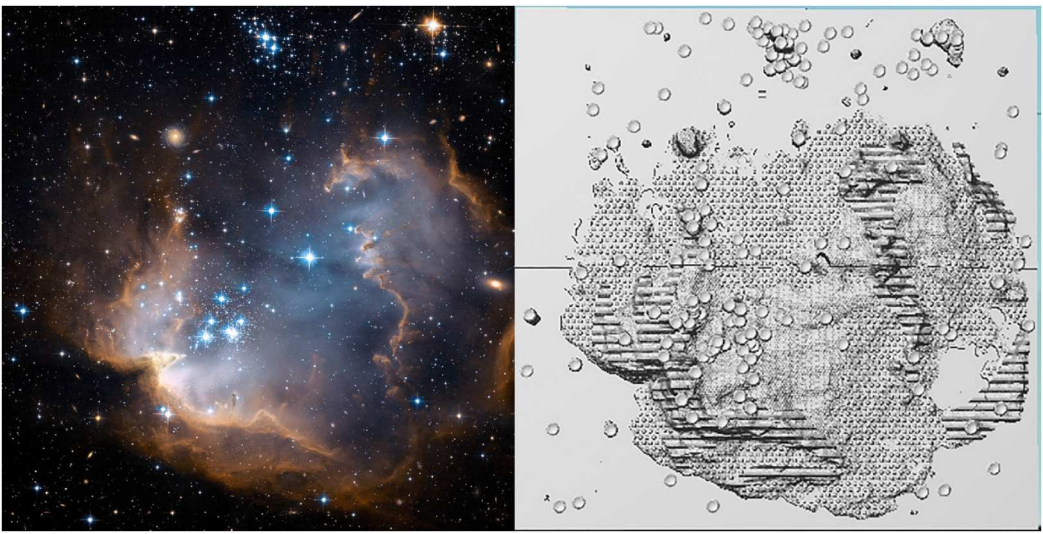

**Figure 3.** NGC602 star cluster HST image (left) and tactile print image (right) used for introducing textures (Carlson et al., 2007; 3DAstronomyPrints, n.d.). The distance to this cluster is 196 thousand light-years, the diameter about 97 light-years with an age of about 5 million years. Image credit: NASA and ESA.

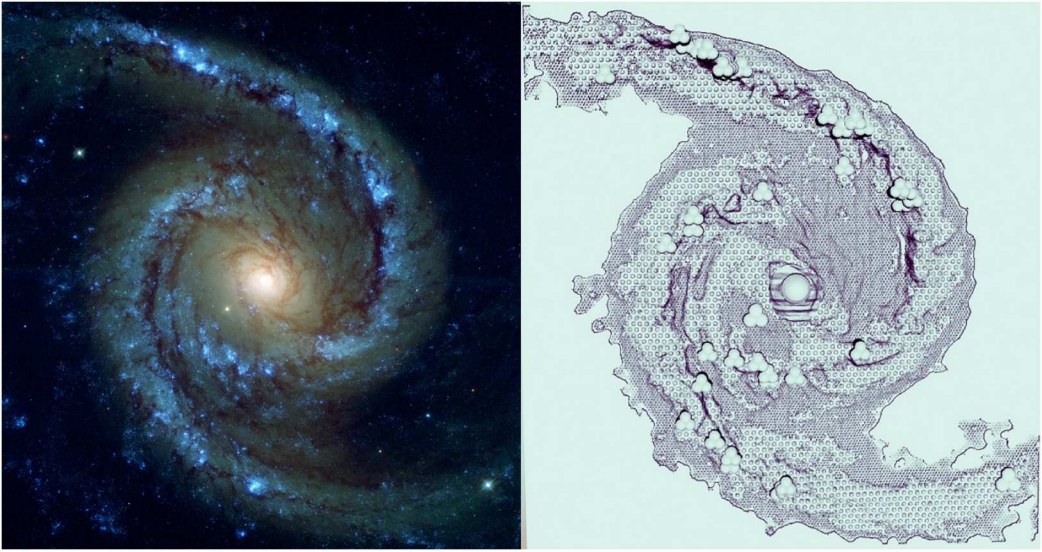

**Figure 4.** Left: HST image of NGC 1566, or the 'Spanish Dancer' Galaxy with distinct spiral arms and a slight bar feature across the central region. The bright core signals material near a supermassive black hole. Bright blue regions signify star formation regions and star clusters. The dark areas are dust clouds and the glowing material is gas. Right: D tactile galaxy representation using our adopted standard textures (Shabani, et al, 2018; 3DAstronomyPrints, n.d.). The distance to this galaxy is 34 million light-years and the diameter about 100 thousand light-years. Image credit: NASA and ESA.

## Methods

### Study overview

This paper describes the final phase of testing for our uniquely textured astronomical 3D prints derived from HST data adopted for both star clusters and galaxies. Previous publications (Christian et al., 2014; Christian et al., 2015; Grice et al., 2015) detailed earlier stages of development and testing of textures and textured 3D prints which involved interactions with volunteers at conferences and workshops, with participants ranging in age from 10 to 82, varied visual acuity, and varied exposure to science. Preferred textures were then applied to an HST image and subjected to more assessment resulting in a consensus recommendation to adopt specific textures as a standard.

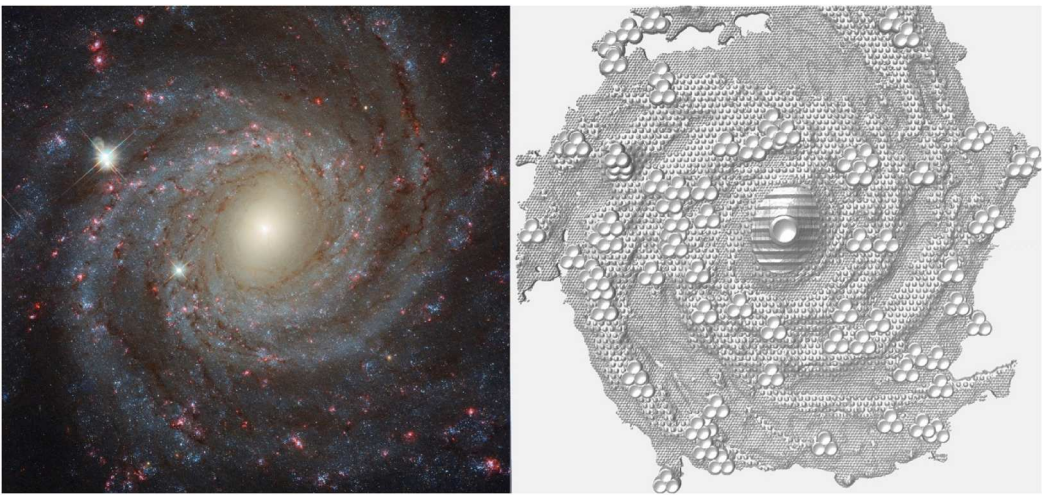

**Figure 5.** Left: HST image of NGC3344. Right: Tactile image of the galaxy. This galaxy was used to assess whether volunteers with B/VI could identify the features and understand that the spiral arms are more tightly wound and fragmented compared with NGC1566 (ESA3344, 2018; 3DAstronomyPrints, n.d.). The distance to this galaxy is 20 million light-years and the diameter about 73 thousand light-years. Image credit: NASA and ESA.

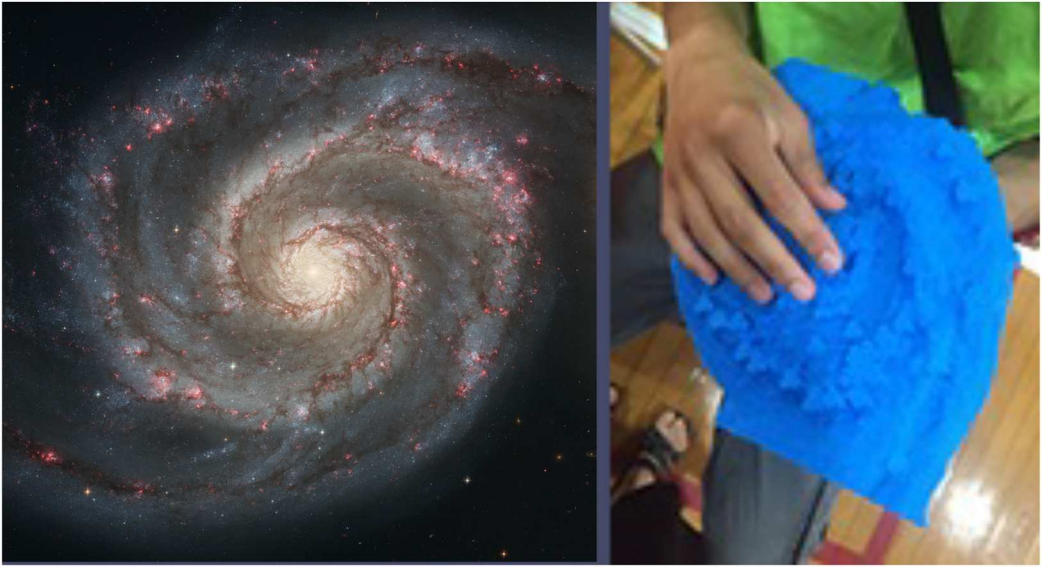

**Figure 6.** The Whirlpool Galaxy. Left: HST image of the Whirlpool galaxy also called M51. Right: Tactile image of the galaxy. This galaxy has more tightly wound arms than NGC1566 but is a classic 'grand design' spiral galaxy with clearly delineated spiral arms (HubblesiteM51, 2005; 3DAstronomyPrints, n.d.). The distance to this galaxy is 23.5 million light-years and the diameter 77 thousand light-years. Image credit: NASA and ESA.

This paper discusses the final project phase assessing the usability and suitability of the standardized textures applied to star cluster and galaxy data. Diverse groups of volunteers participated by exploring the 3D prints and providing feedback on their impressions and reactions to the prints. We also probed whether the testers could translate textures across the suite of prints, discerning shapes, and similarities or differences between celestial object prints.

The testing focused solely on the tactile 3D prints, to avoid the potential influence of ancillary learning materials on assessment outcomes (Black & Wiliam, 1998), thus we explicitly sought simple qualitative reactions to the stand-alone prints. Volunteers received only a brief introduction to the celestial objects represented by the prints. One exception was a group that examined the star clusters first and then the galaxies, which necessitated some description of both types of objects since the physical cosmic scales are so different. QR codes linking to NASA and Hubble websites were provided to all participants to explore if desired.

### Procedure

The qualitative useability testing techniques enhanced our prior methods for informal 3D print evaluation (Christian et al., 2015; Grice et al., 2015). These methods had been informed by our experience evaluating museum exhibits and educational materials based on NASA HST data, along with NASA's educational assessment methods (Christian et al., 2001; NASAEval, 2008; National Research Council, 2015). The usability processes prescribe observations of user insights during open-ended activities and users' collective experiences, thoughts, and feelings while using and discussing a product.

Host institutions interested in our project's potential invited us to events serving individuals with B/VI. Circumstances of testing in informal and potentially noisy environments necessitated adoption of techniques commonly used by museums for interactive exhibits and prototyping in demanding locations (Walhimer, 2012). We also used guidance from the Exploratorium usability testing process of interplay between developer/team in interactions with users described in detail by Socolovsky (2012). Gammon (2003) also outlines useful indicators of interest in museum exhibits.

Interactions involved facilitated dialogues with volunteers to assess content comprehension, gauge curiosity, and monitor conversations and questions arising during exploration. We also observed their mutual discussions of relevant knowledge and experiences, and their engagement with the prints. We documented whether the 3D prints stimulated thought and curiosity, and if the print textures were discernible on each print, either spontaneously or with some coaching. We probed to understand if the testers could draw connections between textures across different celestial objects, enabling them to identify similarities and contrasts. We also engaged in the discussion of the print design and possible customization and adaptations that might be appropriate for different audiences.

### Sample

Usability testing was conducted at several locations with varying demographics and settings, some quite lively and unformatted. We designated two broad categories: structured groups and informal groups. We were familiar with many organizations through conferences or by referrals and their personnel invited us to events at their venues, and specified a time and space for our testing.

### Structured groups

Maryland School for the Blind (MDSB) High School and Middle School after-school program and South Carolina Commission for the Blind (SCCB) STEM Camp catered to students of similar age groups in an informal, out-of-school environment held in designated activity rooms. We expressed interest in middle and high school students as an important and appropriate audience for the subject matter. Both host organizations invited us to attend already organized student events. MDSC regularly conducted a few hours of after-school enrichment activities. SCCB conducted summer camps to introduce students to life skills, physical education, competitions and sports, and some more academic subjects.

### Informal groups

National Federation of the Blind (NFB) held STEM summer events for middle to high school students. Two other public events attracted individuals of diverse ages and backgrounds. One at the National Air and Space Museum (NASM) comprised visitors to the museum ranging from young children to elderly adults. A second venue at a Vision Rehabilitate & Assistive Technology Expo (VRATE) STEM Resources Fair, involved adult individuals interested in technologies and resources for individuals with B/VI. All three of these settings involved high visitor traffic and an open, unregulated flow, making it challenging to track individual comments or collect precise statistics.

### Adaptive usability testing

We used lightly moderated usability testing (c.f., UX Research Team, 2022 for definitions), tailored to the unique characteristics of each location (e.g. Linkedin Community, 2024; California Department of Health, 2017; Socolovsky, 2012). In structured groups, we employed focused interactions with student groups, including time for questions, answers, and general discussion. In the fast-paced informal groups and public forums, we capitalized on brief, opportunistic interactions with individuals as they explored the prints. Instead of the ability to collect detailed, individual feedback, we outlined the project's experimental nature and purpose which fostered open discussion resulting in volunteers providing candid feedback, even when encountering print elements that were confusing or difficult to understand. Volunteers were pleased to participate in active research.

An outline was sketched for gathering responses. Components included introducing ourselves and the purpose of testing the 3D prints which individuals were free to examine and give feedback or not. We clarified that we wished to learn about their reactions, questions, and ideas, not conduct knowledge or skill tests. We asked if the testers minded describing their vision status and if they were familiar with Braille or other tactile products. While individuals then handled prints, we also noted their general knowledge regarding stars, the Sun, and galaxies, and introduced simple astronomical vocabulary. Individuals volunteered information about prior exposure to astronomy, astronomy classes taken, material read, or long-term interest but no formal training, and other feedback. Every volunteer expressed interest and sometimes deep fascination with astronomy which stimulated further discussion about the prints and science. Many individuals offered new ideas, customization of the prints, and adaptations to various circumstances of interest to them personally (home schooling, education for neurodiverse individuals and other ideas).

### *Instrument*

This study generally employed three facets of usability testing (Babich, 2021; UX Research Team, 2022), a common method used to evaluate websites and physical materials before widespread release. First, during the initial general informal interactions and introductions, we asked about vision status and familiarity with assistive technologies and tactile materials.

Secondly, we provided the prints without verbal details to observe volunteer's reactions (c.f., Gammon, 2003): behavior, posture, expression, and verbal feedback indicating interest through active exploration, confident examination, focused attention and curiosity, and apparent comprehension. As testers handled prints, we then used moderated qualitative useability assessment survey questions as a guide similar to prototype usability practices and applications (Socolovsky, 2012). We explored the participants' understanding of the textures, texture distinguishability for each of the features (stars, gas, dust, filament) and recognition of similar textures across different celestial objects. We asked about the identification of shapes, similarities, and differences.

Finally, we engaged the participants in open discussion to explore whether the brief introduction provided at the beginning adequately prepared them to understand the prints and purpose and how much more information they sought. We also asked for ideas on modifications or adaptations that might improve the prints (e.g. Socolovsky, 2012).

The testing outline and questions are in Table 1 in the Supplemental Material. In the organized groups of students with an educator or staff member present, it was straightforward to do the testing. In the unstructured groups, the interactions were short and more ad hoc, requiring *hallway,* or *guerrilla* usability testing procedures (Babich, 2021; UserTesting, 2023). which are designed to last only a few minutes (5–10). This was necessitated by the chaotic, noisy, and fast-paced interactions with the volunteers and availability of only one or two team members.

### *Observations and interactions*

**Structured Groups**. Eight high school and 30 middle school MDSB students participated in 2 h of interaction with our team, examining only galaxy prints, which were in much larger supply and due to time constraint. There was a break between the two group sessions. High school students participated first and were more reserved and cautious in the interactions and curiosity took time to build, while the middle school students were lively, candid, and openly curious and interacted with the team and chatted with each other continuously throughout the sessions.

**Table 1.** Sketch of useability instrument for testing 3D tactile print materials.

| # | Component |
|---|---|
| | *General Background Information Discussion* |
| 1 | If you are comfortable sharing, how would you describe your vision status? |
| | Do you have any previous experience with Braille or tactile materials? |
| | *3D Print Examination* |
| 2 | *Textures (Star Clusters)* Four textures are used: rough bumps ('stipple') for dust/gas mixture, smaller, soft stipple for gas, lined texture for filamentary structure and circular divots for stars |
| | Can you distinguish between various textures in the 3D models? |
| | Which particular features seem the brightest/most intense? |
| | After gaining familiarity with one cluster print (usually NGC602): Examine the print of Westerlund 2 – can you find the various different features? What do you notice about the stars in the cluster? |
| 3 | *Textures (Galaxies)* Four textures are used: rough bumps ('stipple') for spiral arms, smaller, soft stipple for the gas/dust areas between the spiral arms, lined texture for the central bulge of the galaxy, and 3 circular divots together for star clusters. |
| | Can you distinguish between various textures in the 3D models? |
| 4 | If star cluster prints were examined first: Can you identify the various textures in galaxies based on your experience with the star cluster prints? |
| 5a | Is it easy to understand the structure of the galaxy using the 3D models? Why or why not? |
| 5b | Some of the 3D models use taller bumps for the spiral arms texture. Which 3D models do you prefer? Do the taller bumps make a difference in how easy or hard it is to distinguish between textures? |
| 5c | What is the brightest feature in the galaxy? |
| 6 | Can you identify the star clusters? How would you describe where they are located in the galaxy? |
| 7a | Which models do you find the most useful: Those with no clusters, small clusters, or large clusters? Why? |
| 7b | Can you detect by touch any difference in the brightness (intensity/energy output) of the different features on the 3D models? |
| 7c | Which particular features seem the brightest/most intense? |
| 8 | After gaining familiarity with galaxy print (NGC1566): Examine the print of another galaxy (usually both NGC3344 and the Whirlpool were examined). Can you find the various different features and describe them? |
| | *General Feedback, Ideas, Next Steps* |
| 9a | How can we improve these models? What do you like and dislike? |
| 9b | What other celestial objects would you like to have a 3D print of? |
| 10 | Has working with these models had any effect on your understanding of and/or interest in galaxies or astronomy? |

The high school students and college entry students at SCCB were specifically interested in science. They, like their MDSB counterparts, were reserved at first, but after an interaction period became engaged and curious. Several had a surprising amount of astronomical knowledge (surprising since astronomy is not normally in the school curriculum nor elsewhere at the camp). We had two sessions, one with 12 students and one with 15 on different days. Each group spent a total 3 h examining star clusters and then galaxies with a 15-minute break in between.

All student groups expressed interest in the research aspect of the project, leading to animated discussions about their thoughts and learnings.

**Informal Groups**. The 2017 NFB 'Youth Slam' week-long STEM summer program at NFB headquarters was aimed to engage middle and high school students with B/VI in a variety of STEM activities (e.g. Shaheen, 2011 description of 2011 event). Students participated in NFB hands-on learning opportunities related to STEM curricula, focused topical projects (astronomy, nanotechnology, etc.), model rocket building, and an informal exhibition where students could try various materials, technologies, and resources. Exhibits were provided by NASA, commercial vendors, and private institutions. We were invited to support a booth with the galaxy prints to document student reactions to the materials. Our exhibit was near the front of the exhibit hall, so usually students visited our table before finding any of the other NASA astronomical tactile materials. We directly interacted with 35 individuals while other students wandered past the booth and handled prints declining commentary. Some students were sophisticated users of assistive technologies (braille typewriters, readers, audible translators, etc.) so their mutual discussions were quite technical regarding the prints, printer software, and technologies.

NASM of the Smithsonian Institution invited us to set up a table in the main exhibit hall in a broader astronomical and aeronautical environment. Our table was fortuitously placed in front of the HST full scale 13.2 meters high engineering model which attracted visitors to our area.

One NASM curator took an interest in our project to learn methods to make exhibits accessible to visitors with B/VI. This person encouraged our visit, and observed the proceedings, guiding visitors to us. Visitors had an opportunity to explore a plethora of information about HST, imagery, and the resulting science after they examined the prints. The individuals who examined the prints were casual visitors to the museum, including students on field trips, and were delighted and surprised that they could actually handle materials in the museum (most of the holdings of NASM are not touchable as they are historical artifacts). The age range was from middle school children up to senior citizens and most were sighted. A local group of 7 adults from an organization working towards independent living for individuals with B/VI also came specifically to examine our materials. Individuals visited the table at a rate of 100 per hour for a duration of 3 h which afforded a great opportunity to expose the materials but was challenging to obtain feedback systematically since most one-on-one encounters with our two team members were necessarily brief. We captured commentary between visitors besides our direct interaction with them. Most visitors gave us very clear feedback in response to questions.

The VRATE event attracts adults with vision, visual impairment, or blindness. Attendees visited the exhibit hall where we were invited to display prints on a small table. Thirty individuals examined the prints for varying durations from a few minutes to 20 min or longer; those engaging in the longest discussions had extensive experience with tactile materials. The feedback was very useful due to the high-level familiarity with all kinds of materials and approaches to accommodating visual impairment. We were encouraged to present at other venues that could benefit from exposure to our prints.

### *Recording*

In most cases two team members were present per session using a tag team approach, so a presentation was generally done by one and written note-taking by the other. This method was challenging even in structured groups as qualitative assessment format is intentionally informal. Individuals continually commented on prints and responded to our questions and discussion, challenging our ability to question, respond, and record simultaneously. The basic design of this type of testing was to capture the overall impression of group feedback rather than documenting the detailed responses and reactions of each individual unless they provided particularly unusual comments. After the sessions, the team discussed the written notes and recollections and compiled them according to each group, following the outline. The summaries for each group are in Tables 2 and 3 of the Supplemental Material accompanied by the outline in Table 1.

## Results

Our major findings from usability tests are summarized here for each group.

**Structured groups**. The range of vision for the students varied from high to low or partial vision, with 25% blindness, and greater than 2/3 were familiar with Braille.

90% of students including 100% of students with some vision could distinguish features on prints and easily identify the individual textures. The students at SCCB collaborated in pairs or trios and could identify the features of star clusters depicted by texture and found the main clump of stars in each cluster. They note the difference between the two clusters, NGC602 and Westerlund 2, in that the clump of stars in W2 is isolated from the gas and dust features.

Students could discern detailed galaxy structure although one student with blurred vision at MDSC had more difficulty than others because that individual seemed to be using vision more than touch (holding prints close to the face). Comments from MDSC middle school students, who were much more verbal and excited, included the adjectives 'cool', 'diverse' and 'awesome'. High school students preferred prints with more pronounced textures, but the preference was not strong. Some students liked the rougher feel of the prints, others with smaller hands and fingers thought the rougher textures were unpleasant. Middle schoolers tended to like smaller prints with smaller textures.

**Table 2.** Structured groups: Synopsis of usability tests of 3D prints.

| # | 8 High School Students MD School for the Blind Recorded individual responses | 30 Middle School Students MD School for the Blind Recorded comments asn they were verbalized | SCCB Camps – upper level High School, Transition, First year college. Some prior science interest 12–15 per camp |
|---|---|---|---|
| *General Background Information Discussion* | | | |
| 1 | Range of vision from blindness to low or partial vision, 6/8 familiar with Braille | Range of vision from blindness (most) to low or partial vision, 90% familiar with Braille | Range of vision from blindness (3 per camp) to low or partial vision, 70% familiar with Braille |
| *3D Print Examination* | | | |
| 2 | Clusters not examined (time constraint) | Clusters not examined (time constraint) | 95% could distinguish features on star cluster 3D print without help. Stars symbols are brightest. Could find the main features and noticed many more star symbols in W2. Noticed a clump of stars away from the other material (this is the main cluster clump). |
| 3 | 7/8 could distinguish features on prints. Easy to identify textures on 3D prints, and found textures pronounced and 'feelable'. | 90% could distinguish features on 3D print, those with some vision had no trouble with prints features. | 90% could distinguish features on 3D print, those with some vision had no trouble with prints features. |
| 4 | Clusters not examined (time constraint) | Clusters not examined (time constraint) | Most could find star clusters in galaxies. |
| 5a | Discerning the detailed galaxy structure was easy on 3D prints. Only one with blurred vision had more difficulty (individual was using vision more than touch) | Discerning the detailed galaxy structure was easy. 'cool', 'diverse' and 'awesome' | Discerning the detailed galaxy structure was easy on 3D prints. |
| 5b | No preference for larger or smaller textures as long as they could distinguish each type of feature. One liked rougher, another preferred smaller stipple. | Had a hard time telling different stipple patterns on less prominent texture prints. | Rougher texture more distinguishable but less pleasant than smaller texture. |
| 5c | Some had trouble comparing galaxy center to edge. 50% correctly identified galaxy bulges as brightest feature | Easily identified the galaxy central bulge. | Most guessed that the center of the galaxy was the brightest (most had some prior notion of this from news items, etc. on black holes and guessed its identification). |
| 6 | 50% understood the star cluster symbol. Traced arms. | Nearly all could identify star cluster symbol. Traced arms. | All identified the clusters since they were familiar with the star cluster prints and easily traced spiral structure. |
| 7a | All preferred prints with bigger cluster symbol | All preferred prints with bigger cluster symbol | All preferred prints with bigger cluster symbol |
| 7b | >60% could distinguish. | Most said yes, distinguishable. | Most said yes distinguishable. |
| 7c | Clusters and center are brightest. | Clusters are bright, center brightest. | Clusters are bright, center brightest. |
| 8 | (NGC 1566 vs NGC 3344) could tell 3344 was more confused structure. Could identify Whirlpool spiral arms, distinct but tighter wound than NGC1566. | (NGC 1566 vs NGC 3344) could tell 3344 was more confused structure. Could identify Whirlpool spiral arms, distinct but tighter wound than NGC1566. | (NGC 1566 vs NGC 3344) could tell 3344 was more confused structure. Could identify Whirlpool spiral arms, distinct but tighter wound than NGC1566. |
| *General Comments and Feedback* | | | |
| 9a | Most liked touching the models. Students with any sight at all wanted multi-color images | More distinction of clusters in galaxies. Maybe provide a legend. Students with any sight at all wanted color images. | Most liked the models. Students with any sight at all wanted color images. Students also suggested audio and heat for features. |
| 9b | Sun and solar system, planets. Apollo 11, Shuttle, the Sun, comets | Space shuttle, solar system and various satellites. A black hole | Spacecrafts, planets |
| 10 | Yes, 7/ 8 understood. A few still confused by materials and concepts. | Did not discuss. | Yes, 'cool'. |

Table 3. Informal groups: Synopsis of usability tests of 3D prints.

| National Federation of the Blind Youth Slam ~35 students | National Air and Space Museum ~300 individuals | VRATE ~ 30 visitors to the booth Vision Rehabilitate & Assistive Technology Expo 2016 |
|---|---|---|
| **Background:** The majority of students had some visual impairment, with about 30% with blindness. The students were fairly sophisticated in technology, computing, and science | **Background:** Most visitors to NASM were sighted, with an age range of young children to older adults. None of the visitors we were able to question had Braille experience, except for the 7 adults with B/VI who came specifically to learn about our 3D prints. | **Background:** Individuals were familiar with a wide variety of assistive technologies and Braille. About 40% of the individuals who visited our table had blindness. |
| **3D Print Examination:** Students primarily came in groups and discussed prints with our team and each other. Students with partial sight immediately understood the galaxy materials, distinguished the different textures and compared objects easily. They openly discussed what they knew or remembered about astronomy. Students assisted each other in exploring the galaxy prints. Students debated among themselves about detailed characteristics of the textures such as roughness, pattern (stipple, lines, cylinders). | **3D print Examination:** In dept surveying could not be accomplished due to volume of individuals passing through the gallery. Initially, museum visitors were reluctant to touch the 3D prints because 95% of the objects in NASM were exhibits and artifacts that could not be touched. They did look at the prints and notice the shape and textures. Most individuals already had been exposed to or grasped basic astronomy content quickly since their interest was aeronautics, aerospace, space and science, and additionally familiar with HST. Most individuals, including children, understood the textures and generally what they represented on the 3D prints. Many examined more than one print and could identify similar textures from one print to the other, and chose prints at random to handle. The group with B/VI who came by invitation specifically to see our 3D prints were fascinated by the materials. As a cohesive group, it was easier to discuss with them the purpose of the project, the prints format, and the textures which they quickly understood. They had some experience with tactile materials and were able to compare the prints and discuss them as a group. | **3D Print Examination:** All individuals quickly understood the context and purpose of the 3D models and why they were being tested. All of the visitors understood the textures, and could distinguish textures on both star cluster and galaxy prints. Individuals understood and could relate the star clusters to galaxy structures. Most visitors did not have much familiarity with astronomy in detail, yet they grasped basic concepts quickly. This was especially interesting since these adults had much more world experience and had better mental models to understand relative scales than most of the student groups. (Scale is important in understanding the relationship between star cluster sizes and galaxies as well as other practical environments). Individuals gave specific feedback about particular textures, including softening the gas texture (stipple) to make it more evocative of gas. |
| **General Feedback** Students discussed with the team some interesting but expensive augmentations: make areas of the prints produce audio explanations, illuminate different textures with different colored lights, make textures also have some kind of heat indicating perhaps stars and galaxy centers. Some students wanted printers of their own. **Notes** Some students already had some basic astronomy knowledge and had attended a formal session on X-ray | **General Feedback:** A subset of visitors, roughly 50%, asked many questions and offered suggestions about printing other astronomical objects and satellites as well as other random objects. **Notes:** Ancillary materials were prints we provided, along with the full scale HST static test telescope model and the gallery of HST images as a backdrop. This provided essentially an immersion experience in HST astronomy in that part of the gallery, enhancing the context of the 3D | **General Feedback:** Visitors, who were very familiar with all kinds of technologies, wanted heat and light as well as sound on the models (e.g. make the star clusters warm, etc.) Some attendees engaged in very detailed and technical discussions about astronomy, both observational and theoretical, as well as production of the prints, advice on where to use the 3D prints and other suggestions. Several of the visitors were interested in our expertise with 3D printing for a |

(Continued)

**Table 3.** Continued.

| National Federation of the Blind Youth Slam ~35 students | National Air and Space Museum ~300 individuals | VRATE ~ 30 visitors to the booth Vision Rehabilitate & Assistive Technology Expo 2016 |
| --- | --- | --- |
| astronomy during prior days which included 3D prints. Other NASA materials presenting constellations and solar system objects were available elsewhere, augmenting the interest in astronomy, usually after using the prints. | prints for the visitors. The visitors were generally interested in 3D print technology and all about HST and NASA. | variety of other applications they were involved in. |

Some had trouble comparing the galaxy center to the edge. 50% correctly identified galaxy bulges and centers as brightest features and this was correlated with the student's ability to perceive the detailed brightness (height) of the textures relative to each other, although all students could tell the prints had variable height across the object. All individuals preferred prints with bigger cluster symbols in relation to the other textures. In spite of the fact that the students knew very little about galaxies, stars, or star clusters, they all were quick to identify that the galaxy NGC 1566 has an organized spiral structure while NGC 3344 has a more confused structure. All could identify the Whirlpool galaxy's spiral arms as being distinct but more tightly wound than NGC1566. (See Supplemental Material for information on these galaxies).

In the open comment period, students verbalized that they liked the experience of touching the models. Students with any vision wanted multi-color prints similar to press release images which they viewed at the end of the survey period. Students wanted to make 3D prints of a large variety of objects including solar system planets, our Milky Way galaxy, and orbiting satellites. In total, at least 80% understood the materials and concepts. All students wanted more information about star clusters and galaxies. It seemed clear in comparison to the sessions at SCCB and NFB, if time had allowed, the students at MDSC would have benefitted and enjoyed an introduction to the star cluster 3D prints. They also might have understood the nature of the spiral galaxies more clearly even though we were not providing detailed astronomical information to any group until they examined the prints. In structured groups, the most feedback was given from the 30 MDSC middle schoolers who discussed things constantly. The 8 MDSC students were less communicative most likely due to peer pressure and some fear of being embarrassed. The SCCB students gave a substantial amount of feedback and asked many questions. Verbalization came easy to them most likely because they had attended camps together for a week or more and were comfortable with each other as a cohort.

**Informal Groups**. Students at the NFB Youth SLAM events with a range of vision interacted with us in small groups and energetically discussed prints with our team and each other. The sighted students, as with those in the structured groups, understood the galaxy prints more quickly, coached each other in exploring the textures, and chatted about what they knew or remembered about astronomy. Students spontaneously remarked on the characteristics of the textures, clearly distinguishing them and accurately describing them. They freely debated the details of the textures among themselves, and asked the team many questions.

Students discussed with the team some interesting but expensive augmentations: make areas of the prints produce audio explanations, illuminate different textures with different colored lights, make textures also have some kind of heat depending on the features, for example stars and the centers of galaxies would be warm. As expected, students who had attended astronomy sessions and an introduction to Xray astronomy with 3D prints clearly benefited from that background experience when examining the galaxies. Similar to the SCCB students, the students at NFB had been together for several days and so had a good rapport with each other which facilitated discussion.

The NASM visitors discovered our exhibit table while wandering through the museum and also were drawn to the full-scale HST model in the main gallery. Our table was directly in front of the model. Most visitors had an interest in and exposure to astronomy. Their general interest was space, aeronautics, and science, and once they heard a little about the prints, they understood what they represented and were very adept at identifying textures and comparing the different 3D prints. They discussed the prints among themselves and showed each other the textures and different objects. They were interested in our useability test, our project in general, and all about 3D printing technology. They suggested a plethora of objects that could be printed and also asked many questions about HST and NASA.

VRATE was an event showcasing assistive resources and technology for the B/VI community. The format includes formal presentations as well as a substantial exposition of all kinds of materials and equipment. Attendees had a full range of blindness, visual impairment and vision. The testing was particularly interesting because the adult attendees had a wide range of technical expertise, practical know-how, world experience, and were familiar with, had used, or had created assistive technologies. The attendees had no difficulty identifying and recognizing textures including brightness representation, roughness, and scale. They easily identified stars in the star clusters as well as the other features and could relate the textures to the components in the galaxy prints. They could describe similarities and differences between prints. They found the brightest features and wanted to know much more about astronomical details such as star formation and galaxy morphology although few of them had more than a casual knowledge of astronomy. They were very interested in our overall project, future plans, and research.

During print examination, VRATE attendees provided the same wish list of features – color, sound, and temperature – as the Youth Slam students. They provided fairly detailed reactions to the textures such as roughness, distinguishability, and general sensation. They were interested in 3D print technology and how they could employ it for their own projects and personal use.

## Analysis

We compiled written records of all group sessions, organizing the responses according to questions in Table 1. We did not distinguish groups for our main useability testing areas: texture identification, discernability, and translation across prints. First, for texture identification and discernibility, the majority of volunteers could feel and identify different 3D print textures. Regarding translation, the majority of volunteers could identify and find each type of texture on the prints and trace the areas where they occurred. Our secondary usability area was to obtain free-form feedback for future programs. Some volunteers, mostly students, wanted more assistance in understanding the nature of star clusters and galaxies, e.g. such as why stars were clumped together in clusters and why the galaxies appeared to have patterns that were s-shaped or spiral.

Our third aim was to see if any interest in astronomy was stimulated by the experience even without initial supplemental material. Nearly all (more than 90%) of the volunteers asked questions after print examination and wanted to know more about star clusters, galaxies, the universe, planets, and then many other astronomy topics. This result is important for using the materials in future educational and outreach programs. As remarked previously, in informal groups, those with prior access to astronomical information clearly understood the prints before much background was given. They engaged in much more discussion of the prints, what they depicted, and then moved on to discussions of star formation and things they knew or had heard about the universe. Volunteers from all groups provided similar wish lists of augmentations to the prints as described above.

## Discussion

We subjected uniquely textured 3D prints derived from HST astronomical data of both star clusters and galaxies to useability testing as a last phase in our development project. The overarching project aim was to provide a touchable resource as an alternative to visual imagery. Other researchers have also investigated with success that 3D printing is useful for concept development especially in education, for anyone with B/VI. Our results indicated that the testers found that the textures we have developed and the 3D prints we produced were useable, understandable, distinguishable, and a good representation of astronomical data. Also, many volunteers, even those without much exposure to astronomy, perceived the correspondence between the prints for star clusters and those for galaxies and between objects of the same type. The user test groups included students with B/VI who provided robust informal qualitative assessment. Sighted individuals ranging from children to adults also responded favorably to the prints. Adults with B/VI who had experience with various technologies and resources had positive reactions and provided feedback on the prints and in addition, offered ideas and possible avenues to pursue.

### *Beyond the usability test*

We recorded additional observations during the sessions, useful for future print production. Individuals with small fingers (usually younger volunteers) were more sensitive to textures and discerned subtle features and imperfections in the prints. Some were hypersensitive to wider rough stipple texture on bigger prints. Smaller prints are quicker and more economical to produce but not preferred by others such as upper-level high school students and adults. This would be a consideration when creating prints for a specific audience rather than general use. Middle school students had more peer discussions among themselves providing candid observations.

Science-interested high school and older individuals in general grasped the nature of the astronomical objects even without much background in science. Those having more exposure to astronomy provided a lot of additional ideas and feedback for consideration. Discussions of 3D prints among groups of individuals during print examination bolstered collective understanding of astronomy and the prints without our intervention. Sometimes, this circumstance provided more in-depth feedback than structured questioning which is often the situation in museum prototyping and similar environments. Individuals with some sight appreciated that we provided color prints of the objects. Volunteers in informal groups often brought other individuals to our booth or table to examine the 3D prints. In general, volunteers were intrigued by the prints and provided feedback, discussion, and new ideas.

### Conclusions

The development of astronomical textured 3D prints benefited greatly from pilot testing of our 3D prints throughout the many initial stages of the 3D printing project, as documented elsewhere, and the last stage of usability testing in informal environments and group discussions. All these assessments were useful for honing the methods for producing robust 3D astronomical prints with a standard texture suite. The prints preserve the science integrity of the observed data through discernable tactile texturing. Informal interviews with individuals and groups of all ages and visual acuity were able to provide beneficial feedback while increasing their own understanding of astronomy and enthusiasm for science even without much ancillary support materials. As a result, we standardized production of 3D prints for future use in structured programs.

Therefore, this final phase of the production project verified the 3D prints are robust. Based on our past team experience, we know education and public understanding of astronomy are not accomplished with one kind of material. Rather, a full suite of resources is imperative for stimulating engagement, interest, learning and understanding, which was our intent for our future programs.

### *Next steps*

This project demonstrated that 3D printing of astronomical data is a valuable substitute for imagery especially for individuals with B/VI. The 3D prints can integrate into the enormous amount of existing well-vetted information and astronomical background material derived from HST research tuned for public understanding of science and education, representing synergistic approach to maximize the reach of our product.

We provide 3D prints for NASA press releases, (e.g. JPLWR140, 2022; NASA2014-003, 2014) as we intended. Subsequently, we integrated the 3D prints into astronomy summer camps at state bureaus and schools serving students with B/VI (Robinson, 2018) and their teachers in a number of states, augmented with a full suite of other assistive technologies (Madura et al., 2022a; Madura et al., 2022b) to stimulate interest in STEM and STEM careers.

We now regularly produce prints of data from other observatories (for example James Webb Space Telescope, CHANDRA Xray Space Telescope, and ground-based facilities: 3DAstronomyPrints, n.d.) producing prints of multiwavelength data in lieu of imagery.

## Additional teams contributing to the research

The HST Westerlund 2 Science Team
    The HST LEGUS Science Team (https://legus.stsci.edu/team.html)
    The 3DAstronomy@STScI Team (https://tinyurl.com/Astro3D)

## Data sources

Many prints are available through the Astro3D website (3DAstronomyPrints, n.d.)
    and the NASA 3D print repository (NASA3d, n.d.; see Contributors C. Christian and T. Madura).
    All HST data is available through the Mikulski Archive for Space Telescopes, (MAST, n.d.).
    Other NASA data is available through a data repository (NASAData, n.d.)

## Acknowledgements

This work is supported by a contract, NAS5-26555, to the Association of Universities for Research in Astronomy, Inc. for the operation of the Hubble Space Telescope at Space Telescope Science Institute. Innumerable individuals from science museums, universities, commercial companies, and educational institutes provided information, advice, and constructive comments on this project resulting in improvement and integrity for the final 3D print materials.

## Disclosure statement

No potential conflict of interest was reported by the author(s).

## Funding

This work was supported by the Science Mission Directorate [grant number NAS5-26555].

## Ethics

This research has approval through the Johns Hopkins IRB PR00015047.

## ORCID

*Carol A. Christian* http://orcid.org/0000-0002-2179-3308
*Antonella Nota* http://orcid.org/0009-0007-8087-6975
*Noreen Grice* http://orcid.org/0009-0005-3182-2132
*Thomas Madura* http://orcid.org/0000-0001-7697-2955